\documentclass[12pt,final]{iopart} 

 \expandafter\let\csname equation*\endcsname\relax 
 \expandafter\let\csname endequation*\endcsname\relax 
\usepackage{amsmath}
\usepackage{iopams}  
\usepackage{graphicx}
\usepackage{pdfsync}

\begin{document}

\title{Self-assembled Zeeman slower based on spherical permanent magnets} 

\author[cor1]{V. Lebedev}
\address{Department of Physics, University of California, Santa Barbara, California 93106, USA}
\ead{lebedev@physics.ucsb.edu}

\author{D. M. Weld}
\address{Department of Physics, University of California, Santa Barbara, California 93106, USA}

\begin{abstract}
We present a novel type of longitudinal  Zeeman slower. The magnetic field profile is generated by a 3D array of permanent spherical  magnets, which are self-assembled into a stable structure. The simplicity and stability of the design make it quick to assemble and inexpensive. In addition, as with other permanent magnet slowers, no electrical current or water cooling is required. We describe the theory, assembly, and testing of this new design.
\end{abstract}


\maketitle 



\section{Introduction}

Cold atoms have found numerous applications in quantum simulation, quantum chemistry and precision metrology~\cite{phillips98,weiman,bloch,jin,derev,bloom}.  For most such experiments it is helpful to trap as many atoms in a magneto-optical trap (MOT) as possible. In order to achieve this, a collimated slowed atomic beam is often used to load the MOT.  A common approach to atomic beam slowing makes use of the scattering force from a counter-propagating near-resonant laser beam. In a Zeeman slower, the unavoidable Doppler shifts are compensated with Zeeman shifts from an inhomogeneous applied magnetic field~\cite{phillips82}. This field is commonly produced by  a carefully-designed pattern of current-carrying wires around the axis of atomic beam propagation. Recently, alternative designs based on permanent magnets have attracted interest~\cite{ovch07, ovch12,hill12, cheiney11, reinaudi12,hill14}. These use permanent magnets positioned at different distances away from the atomic beam to create the desired field profile. Although the resulting field cannot be switched on and off, the use of permanent magnets offers numerous practical advantages including robustness, ease of maintenance, the elimination of electrical power and water cooling requirements, and low cost. 

Permanent magnet Zeeman slowers can be categorized by the direction of the field relative to the beam axis (longitudinal or transverse) and by the mechanism for positioning the individual magnets. For example, in the work of Hill \emph{et al}~\cite{hill12,hill14}, the positions of the magnets are fixed with screws, which can be manually adjusted for fine optimization of the profile of the magnetic field. Reunaldi \emph{et al}~\cite{reinaudi12} realized an automated version of this design by adding servo motors to control and optimize the positions of the magnets.  Another approach is to position the magnets in a Halbach arrangement~\cite{cheiney11}. All the above-mentioned designs rely on rigid attachment of magnets to a mechanical construction whose position can be fine-tuned if required.

In this paper we present a conceptually novel approach, which relies on ``self-assembly'' of spherical neodymium iron boride (NdFeB) magnets into stable structures. The resultant slower is of the longitudinal-field type, meaning that a magnetic field along the atomic beam axis is used to compensate for the Doppler shift of the atoms. Unlike  transverse-field slowers, which use linearly polarized ($\sigma^{+}+\sigma^{-}$) cooling light, longitudinal slowers use only $\sigma^{+}$ or only $\sigma^{-}$ polarized light. As a result we expect our slower to require roughly half the laser power of transverse-field permanent-magnet slowers.  The design does not require any supporting mechanical holders, and is stable, quick to assemble, and inexpensive.

In section~\ref{design} we present the general approach and details of the main design considerations. In section~\ref{realization} we describe the construction of the slower.  In section~\ref{results} we discuss the measured field profile and calculated performance, and we conclude in section~\ref{discussion}.  


\section{Design Considerations}
\label{design}
The basic idea behind our design is to create a longitudinal-field permanent-magnet slower by arranging a large number of individual magnetized spheres in a magnetically stable configuration.  Design goals include simplicity, low cost, ease of assembly, and tunability (which can be achieved by adding or subtracting individual magnetic elements).  Previous work has shown that an array of  magnetic dipoles can be rigidly positioned so as to create a spatially varying longitudinal field which is suitable for a Zeeman slower~\cite{ovch12}.  In this paper we concentrate on finding an arrangement of magnetic dipoles which both locally minimizes the magnetic interaction energy among the dipoles and creates a suitable slowing field. We refer informally to such an arrangement as a ``self-assembled Zeeman-slower.'' 

\begin{figure}[htbp]
\centering
\includegraphics[width=10cm]{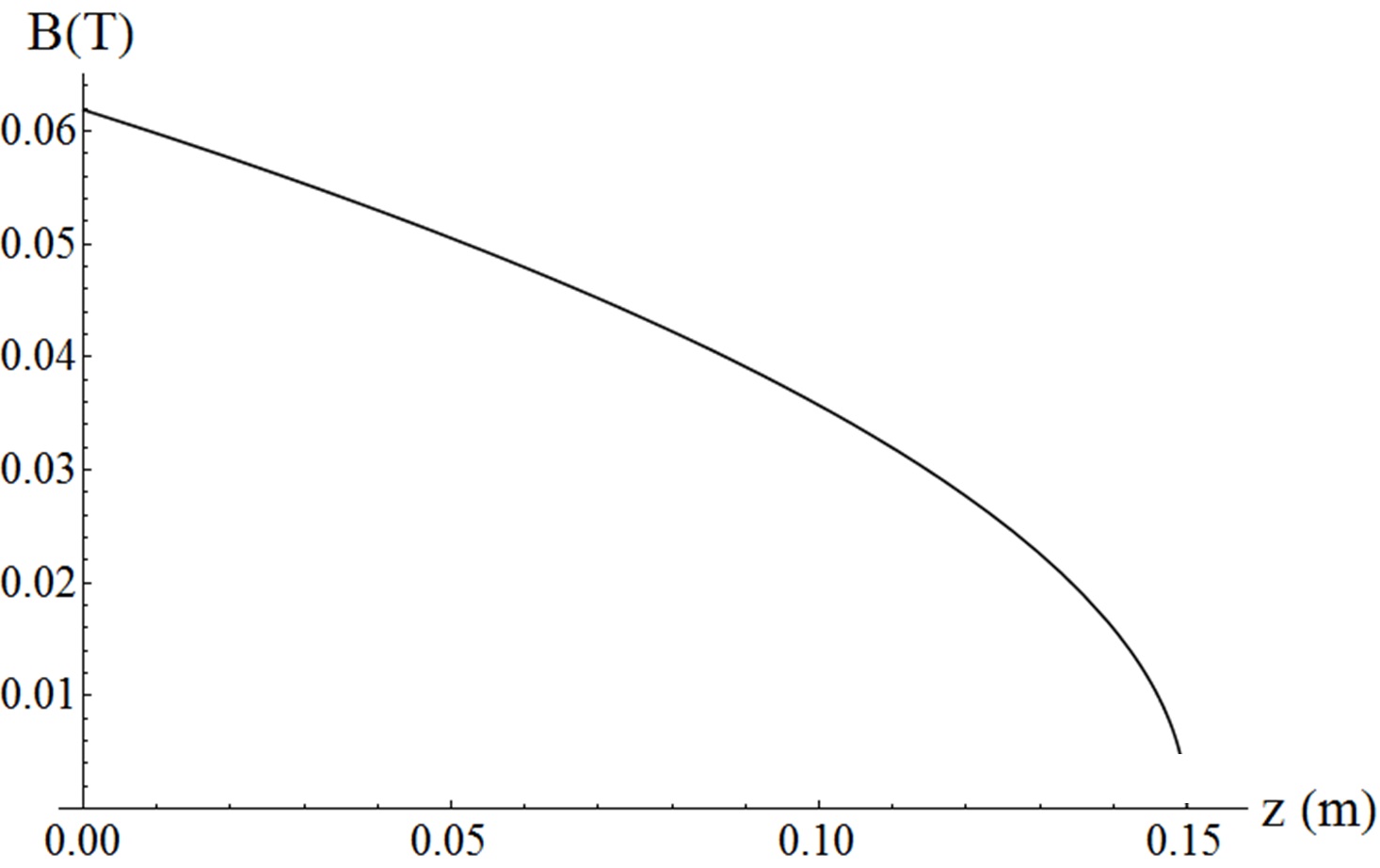}
\caption{Ideal magnetic field profile for Sr atoms based on equation (2), with $L=15~cm$ and $v_0=400~m/s$. }
\label{b_ideal}
\end{figure}

In a Zeeman slower, atoms are slowed by an approximately constant deceleration force due to photon scattering from a counterpropagating beam.  In the limit of large laser intensity, the maximum deceleration  is $a_\mathrm{max}=F_\mathrm{max}/m=\hbar k \Gamma/2m$, where $k$ is the wavevector of the slowing light, $\Gamma$ is the natural linewidth of the relevant optical transition, and $m$ is the atomic mass.  Robust slowing requires operation at some smaller acceleration $a=\eta a_\mathrm{max}$, where $0<\eta<1$.  For constant acceleration, the velocity as a function of position is then
\begin{equation}
v(z)=v_0\sqrt{1-\frac{z}{L}},
\end{equation}
where $v_0$ is the initial velocity, $L=v_0^2/2\eta a_\mathrm{max}$ is the length of the slower, and $z$ is the position along the axis of the slower.  To maintain resonance, the Doppler shift arising from this variation in velocity must be compensated by the Zeeman shift due to an applied magnetic field.  For a transition with a magnetic moment of one Bohr magneton, the condition for such compensation is $kv(z)=\pm \mu_B B(z)/\hbar$, leading to an ideal magnetic field profile of the form
\begin{equation}
B(z)=\pm\frac{\hbar k}{\mu_B}v_0\sqrt{1-\frac{z}{L}}.
\end{equation}
Figure~\ref{b_ideal} illustrates a magnetic field profile calculated from equation (2) with $L=15~$cm and $v_0=400~$m/s. 

In our Zeeman slower design, a field of the appropriate form is created by a superposition of fields due to individual elements.  The components of the field due to a single magnetic dipole placed at the origin and oriented along the $z$-axis are
\begin{equation}
\begin{aligned}
&B_{x} = \frac{\mu _{0}M}{4\pi} \left[ \frac{3xz}{r^5}  \right],\\
&B_{y} = \frac{\mu _{0}M}{4\pi} \left[ \frac{3yz}{r^5}  \right],\\
&B_{z} = \frac{\mu _{0}M}{4\pi} \left[ \frac{2z^2-x^2-y^2}{r^5}  \right],
\end{aligned}
\end{equation}
where $M$ is the magnetic moment of the dipole and $r=(x^2+y^2+z^2)^{1/2}$.  The field of our slower is a sum of such terms due to dipoles at different positions.  For an array of $N$ magnets which has discrete or continuous rotational symmetry about the $z$ axis, $B_{x}$ and $B_{y}$  are zero at all points on the axis of symmetry, and the total magnetic field on axis is the sum of the $B_{z}$ components due to each individual magnet:

\begin{equation}
 B_{z_{~total}} = \sum_{i=0}^{N} B_{z_{i}}~.
\end{equation}

\begin{figure}[here]
\centering
\includegraphics[width=10cm]{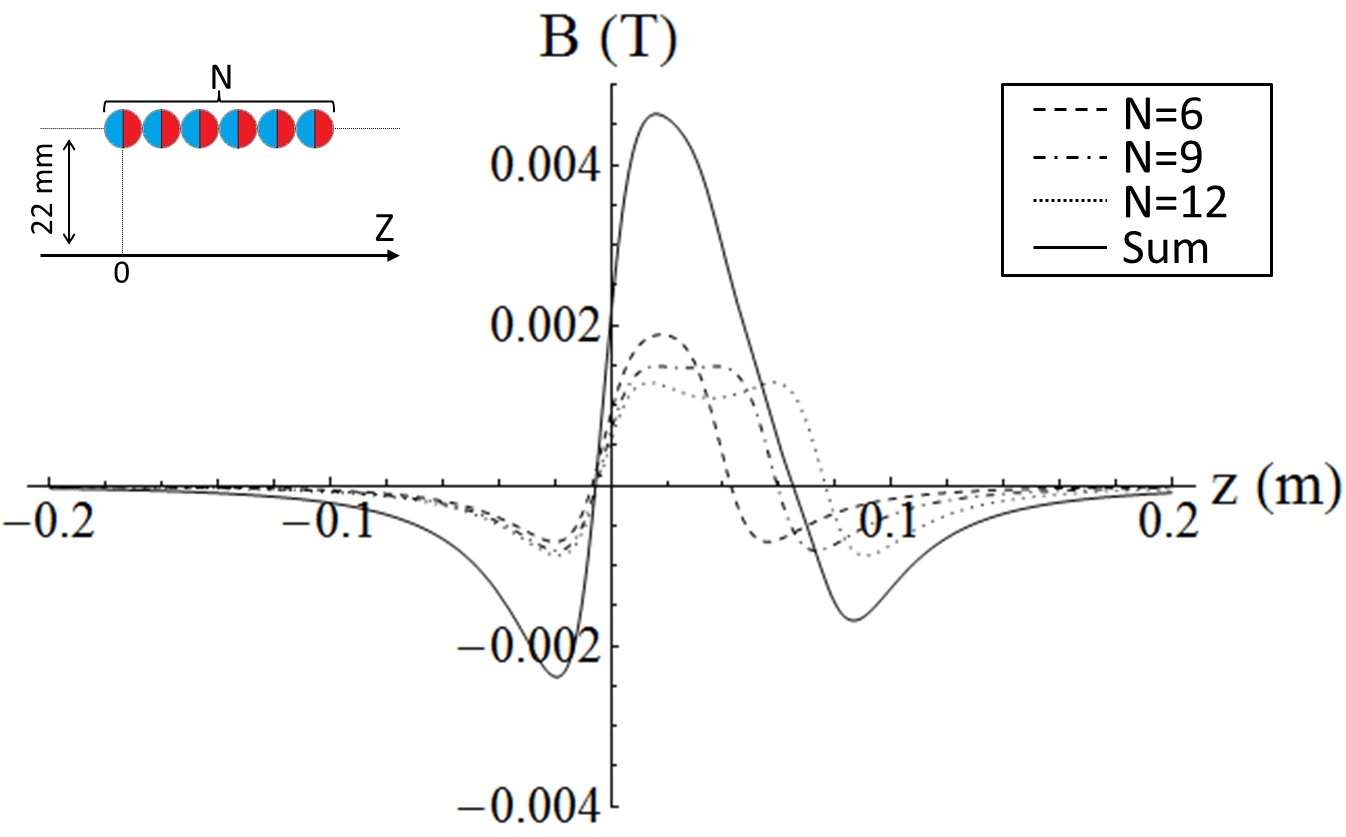}
\caption{Magnetic field from 1D arrays of dipoles.  The left edge of each array is at the origin, and each individual magnet is approximated as a point-like dipole positioned 22 mm away from the $z$-axis and aligned along it. The spacing between neighboring dipoles is 6 mm, and each dipole moment is 0.075 A$\cdot$m$^2$. The dashed line corresponds to an array consisting of six such magnets, the dash-dotted to an array of 9 magnets, and the dotted to an array of 12 magnets. The solid line is the sum of the fields of all three arrays.}
\label{lines2}
\end{figure}

Fig.~\ref{lines2} illustrates the fields from a combination of 1D arrays of different lengths, with each array starting at the origin.  The total field from all three arrays (solid line) has the required asymmetric profile.  It is easy to see that adding the contributions of more such arrays (each with incrementally increasing length) will result in a gradually decreasing total field profile qualitatively resembling that required for a Zeeman slower.  Note that there are two ways to tune the slope of the field profile: by varying the length of each  array and also by varying the distance of each  array from the $z$-axis. The configuration considered in Fig.~\ref{lines2} has all magnets positioned at the same distance away from the $z$-axis.

\subsection{Stable configurations}
For simplicity of construction it is important to ensure that the arrangement of dipoles not only generates the correct field profile but also is magnetically stable.   Although it is possible to fix the position of each dipole with glue or mechanical mounting, here we aim at realizing a stable magnetic configuration which does not require any mechanical support structure.  For optimum beam slowing, $B_{x}$ and $B_{y}$ should be zero and the longitudinal field should vary as little as possible in the transverse direction, which requires the magnet arrangement to be symmetric around the $z$-axis. This means that our structure must be stable in three dimensions.  

\begin{figure}[h]
	\centering
		\includegraphics[width=12cm]{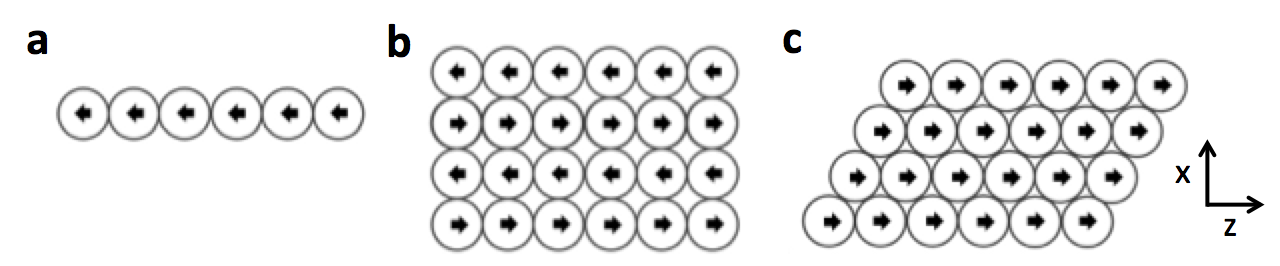}
		\caption{Stable arrangements of magnetic dipoles in 1D (\textbf{a}) and 2D (\textbf{b} \& \textbf{c}). 
		}
	\label{fig:stable_B_0}
\end{figure}

Here we briefly consider some basic examples of structures which are stable in one and two dimensions.  A stable 1D structure is simply a line of  spherical magnets with the south pole of one magnet touching the north pole of the other.  In 2D, stability can be achieved for two different patterns (see Fig.~\ref{fig:stable_B_0}). Note that only one of the solutions corresponds to a non-zero total magnetic field at long distances.


\begin{figure}[h]
	\centering
		\includegraphics[width=9cm]{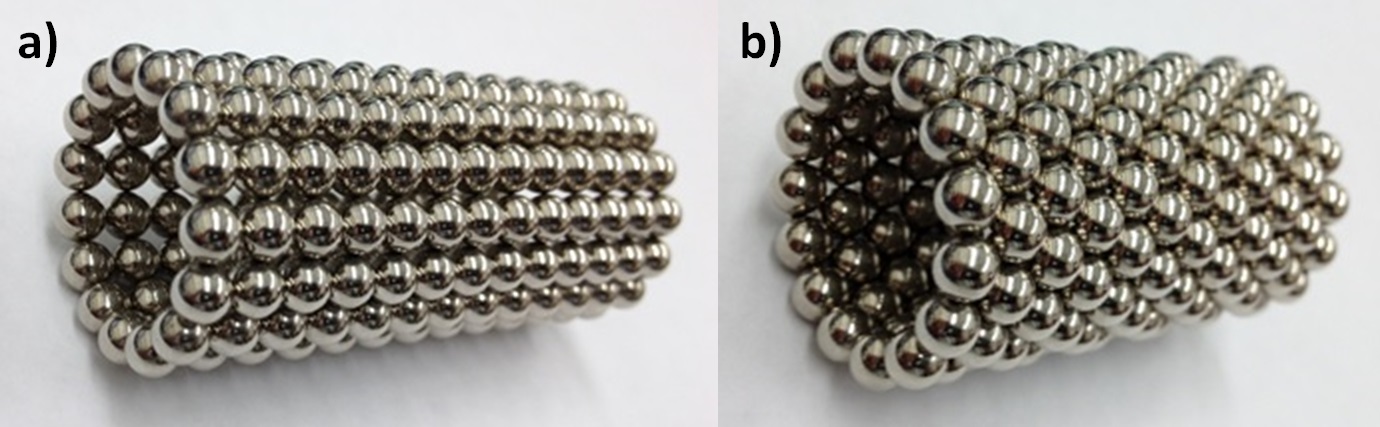}
	\caption{Magnetically stable cylinders with a square (a) and triangular (b) pattern.}
	\label{fig:cylinder_square_pattern}
\end{figure}

The 3D case is more complex than 1D and 2D, and supports an infinite number of stable structures.  As a practical note, stable 3D arrangements can also be very sensitive to the chronological order in which the dipoles are assembled. One aim of this work was to find a 3D pattern that is stable and results in a Zeeman-slower-like profile of the magnetic field.  The line of magnets from Fig.~\ref{fig:stable_B_0}a can be closed onto itself to make a ring which is very stable and generates no net field through its axis.  If one stacks a number of such rings so that their axes overlap the resulting structure will be a cylinder. Depending on the relative orientation of magnetization in neighboring rings one can end up with two possible types of regular cylinder. Fig.~\ref{fig:cylinder_square_pattern}a illustrates a square lattice which results from each ring being magnetized in the opposite direction to the neighboring ring (similar to the pattern described in Fig.~\ref{fig:stable_B_0}b). Fig.~\ref{fig:cylinder_square_pattern}b illustrates a triangular lattice, for which each ring is magnetized in the same direction as the neighboring ones (similar to the pattern described in Fig.~\ref{fig:stable_B_0}c). Both of these cylinders generate no net magnetic field along their axis. 

Although neither of the two cylinders generate any net magnetic field, they play a key role in the design of our Zeeman slower. We use the square-patterned cylinder as a rigid bottom layer for a stable arrangement described in the next section. We refer to the inner cylinder as an \textit{adhesive layer} , because it keeps the magnets in the outer layers fixed in the correct way. The square-patterned cylinder is especially appropriate for supporting line-shaped outer layers aligned  along the length of the cylinder. 

\section{Practical realization}
\label{realization}
As discussed in the previous section, we create stable magnetic structures by adding lines of spherical magnets on top of an 'adhesive' cylindrical inner layer.  In principle, there is a large number of different arrangements of individual magnets that can generate a Zeeman-slower-like field. In choosing the most appropriate structure for a specific application the main constraints arise from the need to keep the whole structure magnetically stable and symmetric around the z-axis, and the need to minimize the total number of individual magnets in order to keep the cost and weight low.

\begin{figure}[h]
	\centering
		\includegraphics[width=9cm]{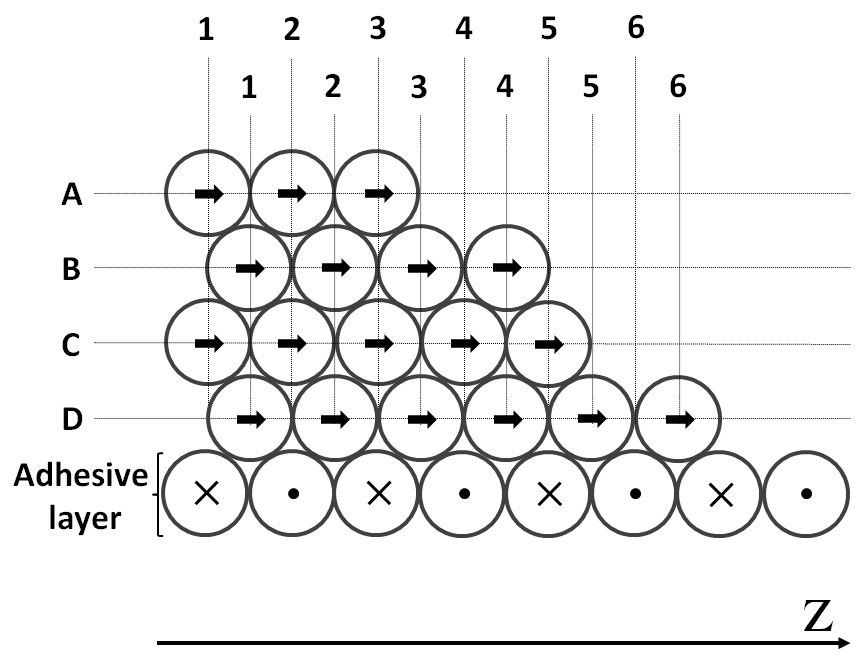}
	\caption{Side view of the structure of each fin.  Apart from the adhesive layer, each magnet is magnetized the same way along the z-axis. Line A is the one furthest away from the axis, while line D attaches the fin to the adhesive layer.}
	\label{fig:fins}
\end{figure}

\begin{figure}[h]
	\centering
		\includegraphics[width=12cm]{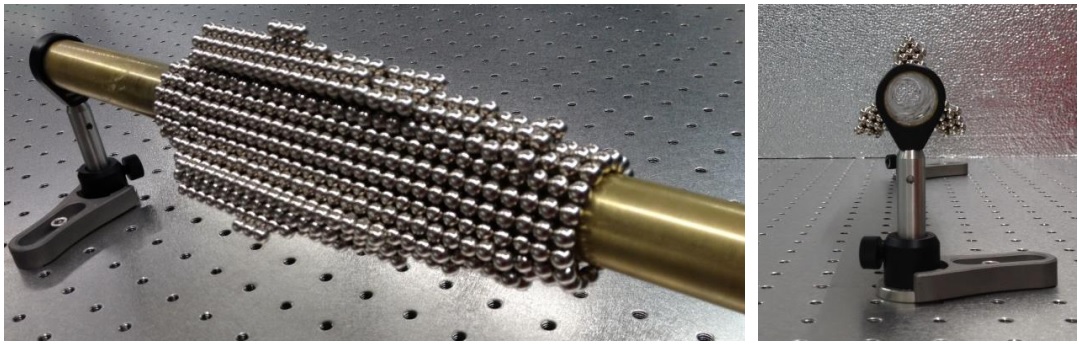}
	\caption{Self-assembled Zeeman slower made of permanent spherical magnets.  \textbf{Left:} side view.  \textbf{Right:} end view.}
	\label{fig:slower}
\end{figure}

For these reasons we chose the shape of the whole structure to have three fin-like elements as shown in Fig.~\ref{fig:slower}.  All three fins are identical and magnetized in the same direction and are held in a fixed position by the adhesive layer, which in turn is wound around a non-magnetic tube. The structure and ordering of magnets within each fin is shown in Fig.~\ref{fig:fins}.  Each individual magnet in the fins is magnetized the same way along the $z$-axis. Adjacent lines of magnets have incrementally different numbers of elements, and each shorter line is incrementally further away from the z-axis. Such a structure results in a slowly decreasing field profile, and three such fins produce a $z$-directed axial field three times larger than than that from any one fin. As seen in Fig.~\ref{fig:slower},  all fins are positioned symmetrically around the z-axis at 120 degrees to each other. Each individual magnet used in this assembly is a 6mm diameter N45 grade spherical neodymium iron boride magnet, with a dipole moment of about 0.075 A$\cdot$m$^2$.

\section{Field Profile and Performance}
\label{results}

\begin{figure}[h]
	\centering
		\includegraphics[width=12cm]{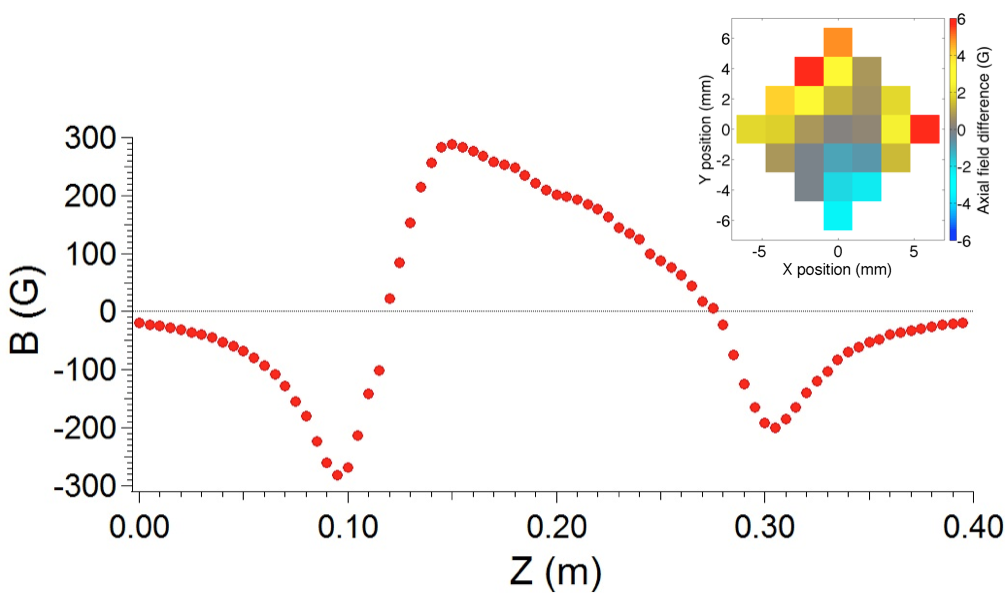}
	\caption{Measured on-axis longitudinal magnetic field profile of the Zeeman slower pictured in Fig.~\ref{fig:slower}.  Inset plots measured variations in the axial field at the maximum field point as the position is scanned transverse to the axis.  }
	\label{fig:slowerfield}
\end{figure}

Fig.~\ref{fig:slowerfield} presents the measured longitudinal magnetic field along the $z$-axis of the slower. The slowing region is from around 15 to 30 cm, resulting in a 15 cm slowing length. This is sufficient for  Zeeman slowing of strontium, due to the broad ($\sim$32 MHz) linewidth and consequent high scattering rate of this atom's 461nm transition. This design could also be straightforwardly adapted to slowing of other atomic species.  The field is reasonably smooth and can be tweaked and improved further by moving the individual magnets around the outer layers of the fins.  Variation of the axial field as the position is scanned in $x$ and $y$ is plotted in the inset of Fig.~\ref{fig:slowerfield}.  These data were taken at the point of maximum field, near the slower entrance.  The variation of the Zeeman shift across a typical atomic beam diameter is small compared to the linewidth of the relevant transition in strontium, indicating that transverse variations in the axial field should not significantly limit slower performance.

\begin{figure}[h]
	\centering
		\includegraphics[width=12cm]{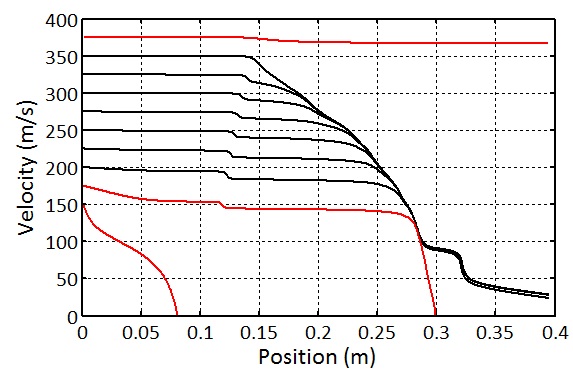}
	\caption{The fate of atoms with different initial velocities. Atomic velocity is plotted as a function of position along the slower axis, for a saturation parameter $s_{0}$ of 1.5 and a detuning $\delta_{0}$ of 340 MHz.}
	\label{fate}
\end{figure}

Using the measured field profile one can integrate the equations of motion to determine the fate of atoms with different incoming velocities. Atoms traveling at velocity $v$  through  a magnetic field $B(z)$ experience both a Doppler shift $kv=2\pi v/\lambda$ and a Zeeman shift $\mu B(z)/\hbar$, where $\lambda$ is the wavelength of the slowing laser beam and $\mu$ is the magnetic moment of the relevant atomic transition. Together with the chosen detuning $\delta_{0}$ of the laser from the atomic transition, the sum of these terms sets the local effective detuning at any point along the $z$ axis:
\begin{equation}
\Delta_{eff}(z) = \delta_{0}+k v-\mu B(z)/\hbar.
\end{equation}
The average spontaneous light pressure force is then given by
\begin{equation}
F(v,z)=\frac{\hbar k \Gamma}{2} \frac{s_{0}(z)}{1+s_{0}(z)+4( \Delta_{eff}^{2}(z)/\Gamma^{2})},
\end{equation}
where $s_{0}(z)$ is the local saturation parameter, which we choose to be constant for simplicity.  The resulting equation of motion can be integrated to find the velocity as a function of time or distance for different starting velocities. Fig.~\ref{fate} presents atomic trajectories for different initial velocities in $(z,v)$ parameter space for the measured field profile of our Zeeman slower. The highest captured velocity is limited to around $370$ m/s, which could be increased by simply adding magnets to increase the maximum magnetic field. Note that the atoms are significantly decelerated in the region between 0.3 and 0.4 meters despite the opposite slope of the magnetic field. This can be explained by the large natural linewidth ($\sim32$~MHz) of strontium's 461nm line, which leads to substantial photon scattering rates even at relatively large effective detunings. Taking this into account, by varying the saturation parameter and laser detuning it is possible to produce a final velocity of around 25 m/s, which is below the capture velocity of a typical MOT. For similar reasons, the atoms with initial velocities below $170$ m/s are decelerated too early and reach zero velocity before reaching the end of the slower. These atoms are therefore lost since they will never reach the MOT region and cannot be trapped.  However,  for a typical oven temperature of  600$^{\circ}$C, the  most probable longitudinal velocity of the atomic beam is close to 500 m/s. This means that there are significantly more atoms with initial velocities between 170 and 370 m/s  than with velocities between 0 and 170 m/s, even before taking into account scattering processes which  can preferentially deplete the population of low-velocity atoms~\cite{hill14}. Therefore this slower design is  capable of trapping more than 75\% of the atoms with starting velocities below 370 m/s. 

\section{Conclusions}
\label{discussion}

We have presented a new type of longitudinal Zeeman slower based on spherical permanent magnets. In particular we demonstrated how one can assemble such magnets into a stable structure which provides an axial magnetic field profile suitable for Zeeman slowing. Like other permanent magnet slowers, this design does not require electrical current or water cooling.  The proposed design is flexible, easy to assemble, and inexpensive.  

\section*{Acknowledgments}
The authors thank Ruwan Senaratne, Shankari Rajagopal, and Zach Geiger for useful discussions and physical insights.  We acknowledge support from the Air Force Office of Scientific Research (award FA9550-12-1-0305) and the Alfred P. Sloan foundation (grant BR2013-110).

\section*{References}


\end{document}